# Label-free Raman spectroscopy and machine learning enables sensitive evaluation of differential response to immunotherapy

Santosh Kumar Paidi[1,#], Joel Rodriguez Troncoso[2,#], Piyush Raj[1], Paola Monterroso Diaz[2], David E. Lee[3], Narasimhan Rajaram[2,^], Ishan Barman[1,4,5,^]

[1]Department of Mechanical Engineering, Johns Hopkins University, Baltimore, MD, 21218
[2]Department of Biomedical Engineering, University of Arkansas, Fayetteville, AR, 72701
[3]Department of Health, Human Performance, and Recreation, University of Arkansas, Fayetteville, AR, 72701
[4]The Russell H. Morgan Department of Radiology and Radiological Science, The Johns Hopkins University School of Medicine, Baltimore, MD, 21205
[5]Department of Oncology, Johns Hopkins University, Baltimore, MD, 21287

[#]Both authors contributed equally to this work.
[^]Corresponding authors: Ishan Barman (ibarman@jhu.edu) and Narasimhan Rajaram (nrajaram@uark.edu)
The authors disclose no potential conflicts of interest.

**Abstract**

Cancer immunotherapy provides durable clinical benefit in only a small fraction of patients, particularly due to a lack of reliable biomarkers for accurate prediction of treatment outcomes and evaluation of response. Here, we demonstrate the first application of label-free Raman spectroscopy for elucidating biochemical changes induced by immunotherapy in the tumor microenvironment. We used CT26 murine colorectal cancer cells to grow tumor xenografts and subjected them to treatment with anti-CTLA-4 and anti-PD-L1 antibodies. Multivariate curve resolution – alternating least squares (MCR-ALS) decomposition of Raman spectral dataset obtained from the treated and control tumors revealed subtle differences in lipid, nucleic acid, and collagen content due to therapy. Our supervised classification analysis using support vector machines and random forests provided excellent prediction accuracies for both immune checkpoint inhibitors and delineated important spectral markers specific to each therapy, consistent with their differential mechanisms of action. Our findings pave the way for in vivo studies of response to immunotherapy in clinical patients using label-free Raman spectroscopy and machine learning.

**Introduction**

The emergence of immune checkpoint inhibitor-based immunotherapy has revolutionized cancer treatment.[1-3] The antitumoral effect of immune checkpoint inhibitors is orchestrated by successfully blocking the receptors on cancer cells (e.g. PD-L1) and tumor-reactive T cells (e.g. CTLA-4 and PD-1) to limit the interactions that inhibit T cell activation and unleash an immune response against cancer cells.[4-6] While immunotherapy has shown incredible responses in certain cancer types such as melanoma and renal cell carcinoma, only a small subset of patients (sometimes as low as 25%) derive clinical benefit from the therapy.[5,7] Therefore, there is an urgent need for predictive biomarkers of response that can identify patients who would benefit from immunotherapy.[8-10] The early determination of non-responders will help them to explore alternative treatment strategies and alleviate the toxic immune-related adverse effects of ineffective therapy.

The current clinical metrics for prediction and evaluation of response to immunotherapy are not very effective.[11,12] For example, the use of immunohistochemistry for PD-L1 expression as a predictive biomarker is highly debated due to the diversity of assays, lack of universal positivity thresholds, and observations of clinical benefit in patients with low levels of expression.[12,13] Similarly, an increase in tumor-infiltrating lymphocytes has been associated with a better response of anti-PD1 therapy in a melanoma study involving a small number of patients.[14] A liquid biopsy strategy combining the blood counts parameters, clinical characteristics, and serum lactate dehydrogenase (LDH) predicted the response of patients without the metastatic disease to anti-PD-1 therapy with about 60% accuracy.[15] In this milieu, several imaging biomarkers are also currently at various stages of development.[16,17] FDG-PET scans of patients treated for melanoma and other cancers provided the evidence of an association between metabolic changes and response to immunotherapy.[17] Several ongoing studies also leverage PD-1/PD-L1 and CTLA-4 targeting antibodies radiolabeled with $^{89}$Z for evaluating the tumor uptake of therapeutics using PET imaging, but they are associated with challenges similar to PD-L1 IHC noted above.[17] Tumor microenvironment composition and metabolism have been shown to play important roles in the regulation of immune response and disease progression.[18,19] While some early studies provided evidence of a correlation between collagen density and immunosuppression, the compositional changes of tumor microenvironment in response to immunotherapy have remained largely unexplored.[20-22] A similar gap in knowledge exists due to the lack of investigations into the overall therapy-induced changes in the lipid composition of the tumors, despite the evidence of changes in lipid metabolism in individual immune cells due to immune checkpoint inhibition.[18,19] Therefore, there is an urgent need for tools that can provide a holistic picture of the evolution of the immune microenvironment with immunotherapy. Such methods are critical to discovering novel biomarkers of response that not only capture the differences in the composition of individual cellular and extracellular constituents but also leverage their complex interactions to achieve unprecedented sensitivity and specificity.

In this study, we propose Raman spectroscopy, a label-free optical technique based on the inelastic scattering of visible-NIR light, to provide an alternate route for evaluating compositional changes in the tumor and its stroma in response to immunotherapy.[23-26] Recently, Raman spectroscopy has emerged as a valuable tool for non-perturbative study of the biological specimen based on light-induced changes in the molecular vibrational modes. The exquisite chemical specificity of this technique has been leveraged by us and others to study a variety of biological applications, including cancer progression and its therapy.[24,25,27-37] Recently, we showed that Raman spectroscopy allows monitoring response to clinically relevant doses of fractionated radiation therapy in mice bearing lung tumor as well as head and neck tumor xenografts.[38] We also showed that pre-treatment Raman spectra were able to classify tumors that were sensitive and resistant to radiation therapy. Several studies at the cell level have also used Raman microscopy to characterize the effects of radiation therapy and chemotherapy on cancer cells. However, despite the success in characterizing the biochemical changes in the tissue in response to cancer progression and radiation therapy, Raman spectroscopy of tumor response to immunotherapy remains unexplored.

Using a CT26 murine model of colorectal cancer and its treatment individually with three doses of anti-CTLA-4 and anti-PD-L1 antibodies, we sought to determine if Raman spectroscopy is sensitive to the changes associated with immune checkpoint inhibitor therapy. We used multivariate curve resolution – alternating least squares (MCR-ALS) decomposition of the spectral dataset and uncover pure component-like spectra that hint at putative biomarkers of response. We showed that the scores of key component spectra that resemble lipid, nucleic acids, and collagen as well as their spatial heterogeneity change significantly with the administration of each immune checkpoint inhibitor therapy. Our leave-one-mouse-out implementation of support vector machines (SVM) established the feasibility for automated classification of independent test tumors according to their response immunotherapy. Finally, we leveraged the feature-ranking capability of bagged random forests to identify the spectral markers of response to each immune checkpoint inhibitor therapy. To the best of our knowledge, the current study is the first Raman spectroscopic investigation of changes in the tumor composition induced by immunotherapy. The results of our pilot study using a portable clinical Raman system show the promise of optical spectroscopy in guiding personalized treatment decisions for patients. Together with our prior success in monitoring response to radiation therapy, our future explorations in this area will attempt to characterize the tumor compositional changes induced by the combinations of radiation therapy and immune checkpoint inhibitors.

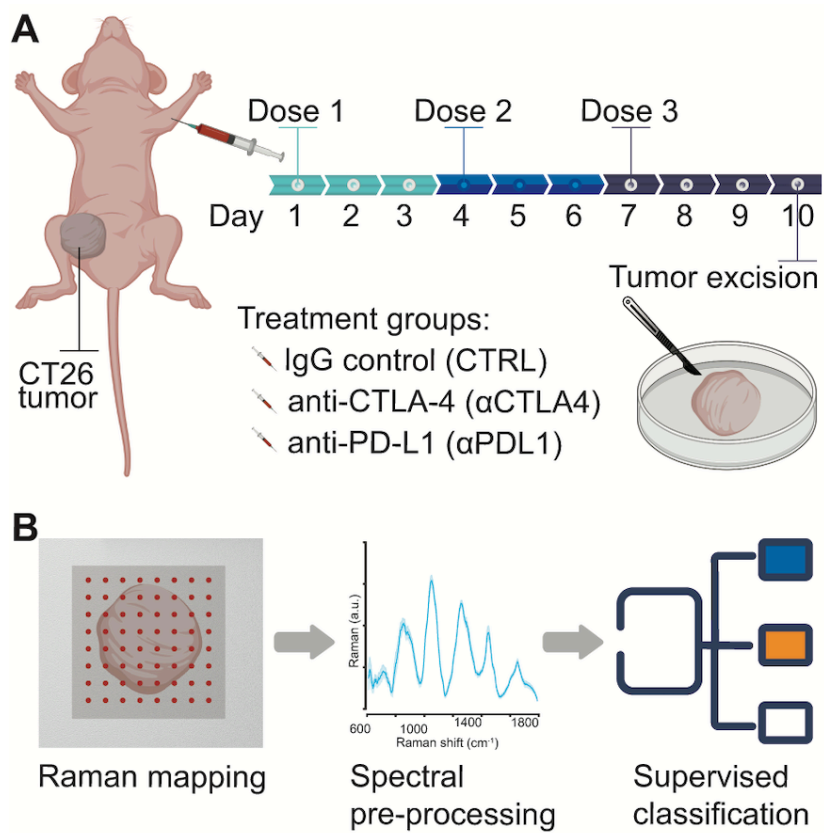

**Figure 1. Label-free Raman spectroscopy for monitoring response to immunotherapy.** (**A**) Study schematic showing the timeline of immune checkpoint inhibitor dosing. (**B**) Overview of the analysis of Raman spectral dataset.

**Results**

***Raman spectroscopy illuminates the biological constituents of tumors***

To evaluate the sensitivity of Raman spectroscopy to the biological response to immunotherapy, we employed murine tumors treated with the immune checkpoint inhibitors anti-CTLA-4 ($\alpha$CTLA4, n = 8) and anti-PD-L1 ($\alpha$PDL1, n = 7) along with control tumors (CTRL, n = 10) treated with IgG antibodies (**Fig. 1A**). The ex vivo Raman mapping of tumors excised after three doses of antibodies in the three groups provided a spectral dataset comprised of 7585 spectra from spatially distinct points (~ 300 on an average) on the 25 tumors (**Fig. 1B**). The mean spectra after background subtraction for each treatment group along with the standard deviation are shown as shaded error plots in **Fig. 2A**. The prominent Raman peaks associated with the biological constituents of the tumors can be observed at 849 cm$^{-1}$ (C-O-C skeletal mode of polysaccharides), 1260 cm$^{-1}$ (amide III of proteins and $CH_2$ in-plane deformation of lipids), 1301 cm$^{-1}$ ($CH_3/CH_2$ twisting or bending modes of lipids and collagen), 1448 cm$^{-1}$ ($CH_2$ bending modes in lipids and collagen), and 1657 cm$^{-1}$ (amide I of proteins and C=C stretching in lipids). However, the average spectra do not present significant macroscopic differences across the three treatment groups. Therefore, we used multivariate curve resolution alternating least squares (MCR-ALS) to decompose the spectral dataset into pure constituent-like component loadings and their contribution (scores) in each spectrum.[39] A five-component MCR-ALS decomposition provided loadings that harbored features of lipids (MC1), nucleic acids (MC2), and collagen (MC3) as shown in **Fig. 2B**. The important features for each component loading are shown in the figure and their band assignments are tabulated in **Table ST1** (supporting information). The remaining two loadings that show formalin fixation and weak collagen features are provided in **Fig S1** (supporting information) along with their assignments in **Table ST1.**

***MCR-ALS analysis reveals significant differences in compositional constituents of spectra***
To evaluate the changes in the biochemical composition of the tumor in response to immunotherapy, we compared the MCR scores of the pure component-like loadings for spectra in the three treatment groups. As shown in **Fig. 2C-F**, we performed three comparisons for each component to study the effect of immune checkpoint inhibitors (CTRL vs αCTLA4, CTRL vs αPDL1) on the distribution of their scores and to elucidate the differences between the two therapeutics (αCTLA4 vs αPDL1). The violin plots (with embedded box and whisker plots) in **Fig. 2C** shows that there is a slight but significant increase in the scores of lipid-like MC1 component in the αCTLA4 treatment group compared to the CTRL. However, there is a significant decrease in the lipid scores due to αPDL1 treatment. The scores of the nucleic acid-like MC2 component, however, decreased significantly for both the treatment groups in comparison to the control group CTRL and the effect was more pronounced for the αPDL1 group (**Fig. 2D**). In contrast to MC1 and MC2 scores that showed larger effect sizes in the αPDL1 group, the scores of collagen-like MC3 component reduced significantly and with a larger effect size for treatment with αCTLA4, while they did not change significantly with αPDL1 treatment (**Fig. 2E**). The variation in the effect sizes (**Fig. 2F**) shows the differential perturbation of the tumor composition upon treatment with αCTLA4 and αPDL1.

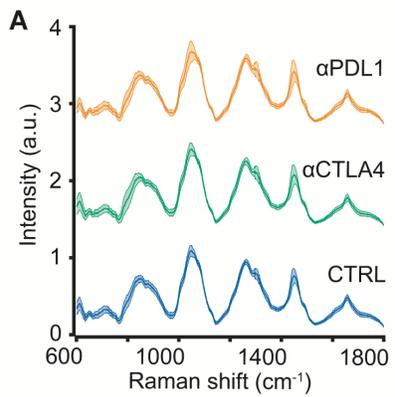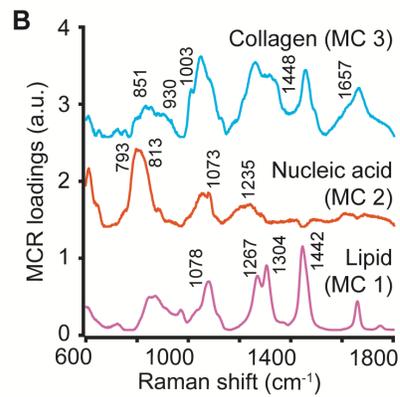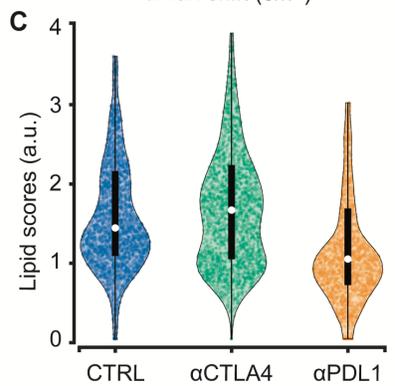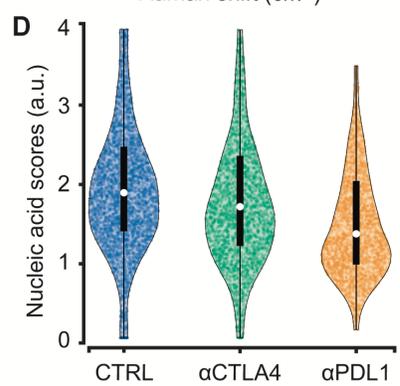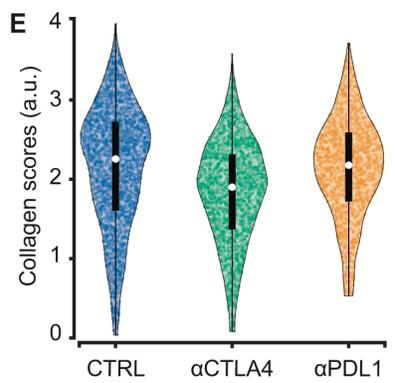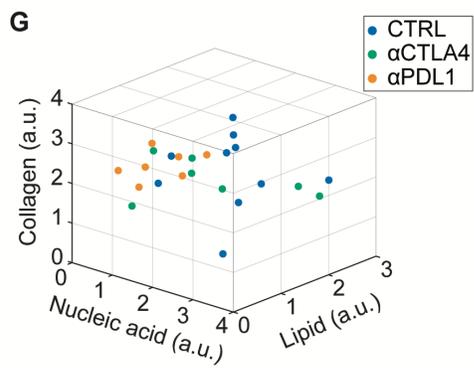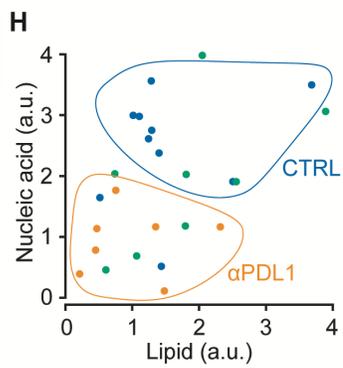

**Figure 2. Raman spectroscopy of response to immunotherapy.** (**A**) The shaded error plots show the mean and standard deviation of the Raman spectra collected from the tumors treated with three doses of immune checkpoint inhibitors. (**B**) A selection of MCR-ALS components derived from the entire spectral dataset that show features similar to lipids (MC1), nucleic acids (MC2), and collagen (MC3). The comparisons of scores for each of the MCR components resembling lipids (**C**), nucleic acid (**D**), and collagen (**E**) across the four treatment classes are shown. The statistical significance as assessed by Wilcoxon rank-sum test p-value and effect size for each of the three binary comparisons between treatment classes (rows) and three MCR components (columns) are shown in panel (**F**). (**G**) The median MCR scores for each tumor in the three treatment classes are shown as a 3D scatter plot. (**H**) The 2D projection of the plot is presented to show the clustering of CTRL and αPDL1 groups.

The differences in the trend and effect sizes of MCR score distributions across the three classes motivated us to explore the possibility of discriminating the treatment classes from control in the subspace formed by the pure component-like MCR loadings. We used the median MCR scores for each animal to illustrate their relative co-localization in the 3D space defined by MC1, MC2, and MC3 loadings (**Fig. 2G**). We noted that there are no clear class boundaries between the three classes in this subspace despite subtle but significant differences observed in the univariate comparison of the MCR scores for each component loading. However, we examined the 2D projection (**Fig. 2H**) of the 3D data on to the plane defined by MC1 (lipid-like) and MC2 (nucleic acid-like) axes due to the relative clustering of mice in the aPDL1 treatment group and the observed higher effect sizes for MC1 and MC2 scores in the CTRL vs aPDL1 comparison (**Fig. 2F**). We found that there is a weak clustering of the CTRL and aPDL1 mice in this plane with a couple of CTRL mice residing in the relatively tighter aPDL1 cluster, consistent with the observations in univariate analysis of MCR scores. The results along with the absence of a non-overlapping cluster for aCTLA4 group show that the MCR scores alone may be insufficient for assessment of response to immune checkpoint inhibitors.

Next, we sought to quantify the spatial heterogeneity of the MCR scores and assess their variation with immunotherapy. We employed a metric (referred to as heterogeneity index in the current manuscript) developed for measuring spatial heterogeneity in medical images based on the distance-dependent deviation of the spatial distribution from the smoothest intensity gradation.[40] The calculated heterogeneity indices showed significant differences for each component with $\alpha$CTLA4 and $\alpha$PDL1 treatment (**Fig. S2**, supporting information). The spatial heterogeneity of scores of lipid-like and nucleic acid-like MCR loadings decreased with immune checkpoint inhibitor therapy for both the groups. However, for the collagen-like component, we observed a decrease in the spatial heterogeneity of MCR scores for αCTLA4 group and an increase for αPDL1 group. These subtle differences in spatial differences show that Raman spectroscopy can not only capture variations in the compositional constituents of the tumors but also illuminate the evolution of their spatial information in response to therapy.

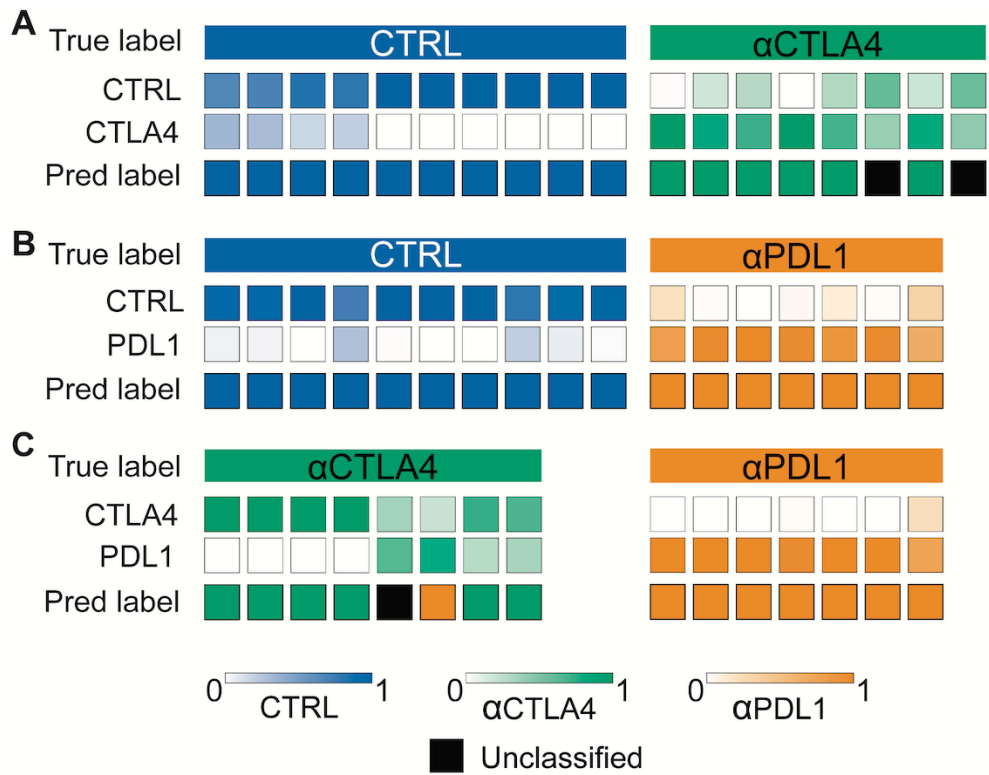

**Figure 3. Supervised classification of response to immunotherapy.** The results for leave-one-mouse-out SVM analysis are shown as heatmaps for the prediction of response to immune checkpoint inhibition therapy for the treatment groups – (**A**) CTLA4 and (**B**) PDL1 in comparisons to the control group CTRL, and (**C**) for evaluating the inter-therapeutic differences between the treatment pairs (αCTLA4 vs αPDL1). The top row in each heatmap shows the true label of the mice (columns of the heatmap) in the group. The central rows in the heatmaps show the distribution of the predicted labels for spectra from each mouse into the classes being compared. The bottom rows show the overall class prediction for each mouse after thresholding on the prediction frequencies.

*Support vector machines allow prediction of response to immune checkpoint inhibitors*

Since the combination of MCR scores did not provide clear class boundaries between the control tumors and tumors treated with $\alpha$CTLA4 and $\alpha$PDL1, we used support vector machines (SVM) to build a supervised classification routine for predicting response to immunotherapy. We employed a leave-one-mouse-out protocol (described in the Methods section) to iteratively use the spectra from each individual mouse as a test dataset to the binary SVM models trained on the spectra from the remaining mice in the dataset. Based on the spectral prediction frequencies obtained for each class, the test mice are either classified correctly, misclassified, or unclassified according to predetermined thresholds described in the Methods section. We performed the three binary comparisons as before to test the ability of SVM to predict the response to therapy using individual immune checkpoint inhibitors and distinguish between the tumors treated with $\alpha$CTLA4 and $\alpha$PDL1 (**Fig. 3**). First, the binary SVM classification between CTRL and $\alpha$CTLA4 yielded accurate predictions for mice belonging to both the groups except two mice in the $\alpha$CTLA4 group that got unclassified (**Fig. 3A**). Next, the classification between CTRL and $\alpha$PDL1 groups resulted in accurate prediction of all the mice in both the classes, consistent with the observation of a weak boundary between the two classes in MCR-ALS analysis (**Fig. 3B**). Finally, the comparison between $\alpha$CTLA4 and $\alpha$PDL1 groups to check the inter-therapeutic differences between the responses provided an accurate prediction of all the mice treated with $\alpha$PDL1 and most of the mice treated with $\alpha$CTLA4 (**Fig. 3C**). However, we observed that one of the eight mice in the $\alpha$CTLA4 cohort got misclassified as $\alpha$PDL1, while another got unclassified.

*Random forest analysis uncovers distinct spectral markers of response to therapy*

The availability of spectral datasets from the CT26 tumors treated with two different immune checkpoint inhibitors $\alpha$CTLA4 and $\alpha$PDL1 allowed us to employ random forest classification analysis for uncovering the spectral markers of their response. We leveraged the ability of random forests to rank the predictors in the order of their contribution to the classification tasks and determined the five most important Raman predictors (wavenumbers) for response to each therapeutic (**Fig. 4**). Since we sought to compare the differential composition of the tumors treated respectively with $\alpha$CTLA4 and $\alpha$PDL1 to CTRL, we chose the binary classification tasks – CTRL vs $\alpha$CTLA4 and CTRL vs $\alpha$PDL1 for this analysis. We first tested the out-of-bag classification error for binary classification tasks using random forest training on the entire spectral dataset with bootstrapped aggregation (bagging). The out-of-bag classification error for both the comparisons decreased to less than 2% upon the inclusion of fewer than 50 trees in the forest (**Fig. 4A**). We also observed that the out-of-bag classification error for both the comparisons followed a similar trajectory with the inclusion of additional trees. Next, we plotted the predictor importance estimates provided by the random forests for each comparison to uncover putative spectral markers of response to immunotherapy. We found that the accuracy of random forest prediction of tumor response to $\alpha$CTLA4 treatment is contributed by the Raman spectral features at or around 981 cm$^{-1}$, 1065 cm$^{-1}$, 1386 cm$^{-1}$, 1574 cm$^{-1}$, and 1652 cm$^{-1}$ (**Fig. 4B**). The same analysis for the tumors in the $\alpha$PDL1 group revealed the spectral features around 1323 cm$^{-1}$, 1574 cm$^{-1}$, 1596 cm$^{-1}$, 1652 cm$^{-1}$, and 1763 cm$^{-1}$ as the key markers of therapeutic response (**Fig. 4C**). We observed that while the two sets of spectral markers had two common features around 1574 cm$^{-1}$ and 1652 cm$^{-1}$, the remaining features were unique to the immune checkpoint inhibitors used for the treatment (**Fig. 4D**). The band assignments for the identified predictors in both the treatment classes are tabulated in **Table ST2** (supporting information).

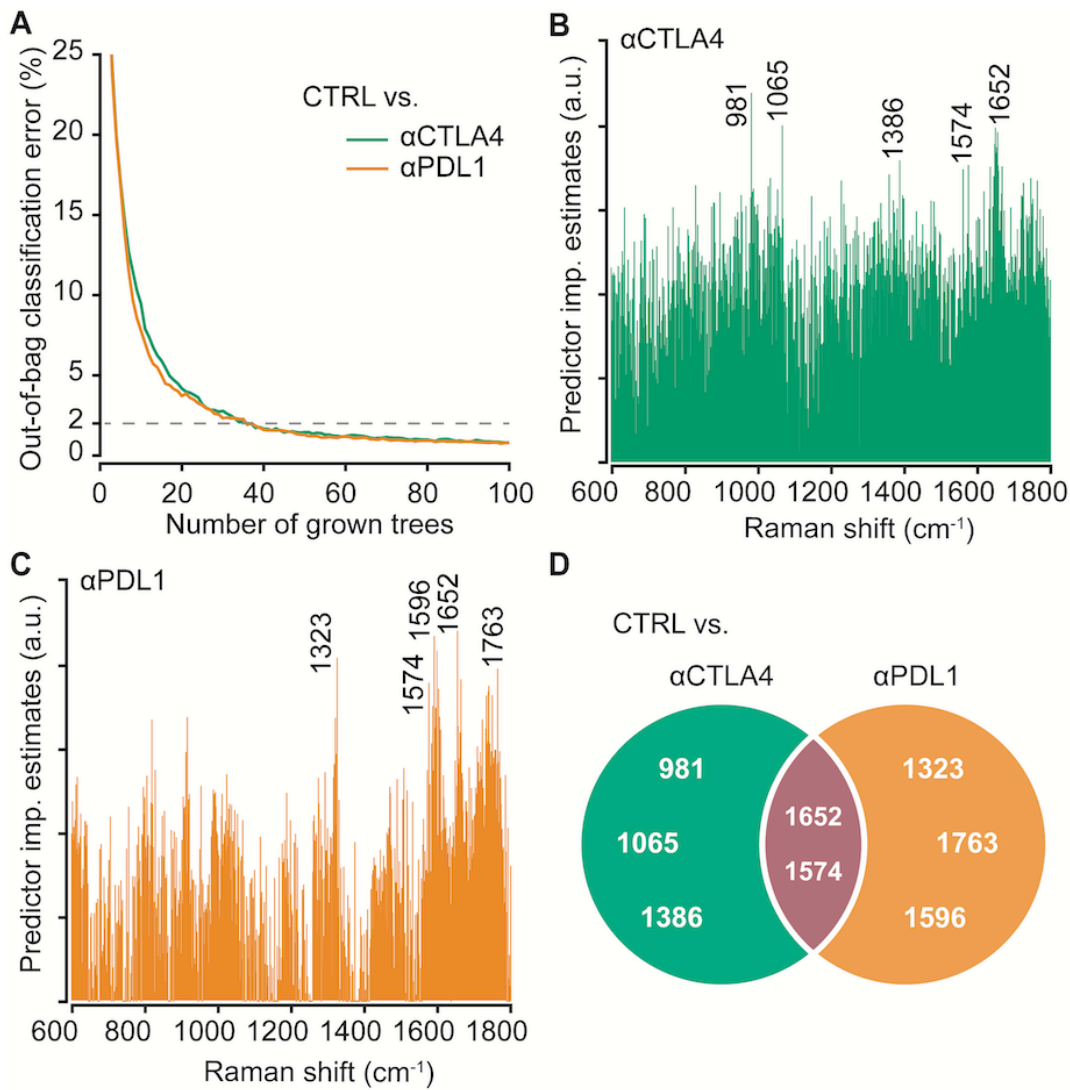

**Figure 4. Spectral markers of response to immunotherapy using Random forest classification.** (**A**) The out-of-bag classification error profiles for bagged random forests trained on the spectral datasets to predict the response to each therapy – CTRL vs αCTLA4 and CTRL vs αPDL1 – are shown. The dotted line indicates a 2% error level. The bar plots show the predictor importance estimates for random forest predictions of response to (**B**) αCTLA4 and (**C**) αPDL1 treatment. (**D**) The 5 most important (non-neighboring) Raman features (cm$^{-1}$) for the prediction of response to each treatment are presented as a Venn diagram.

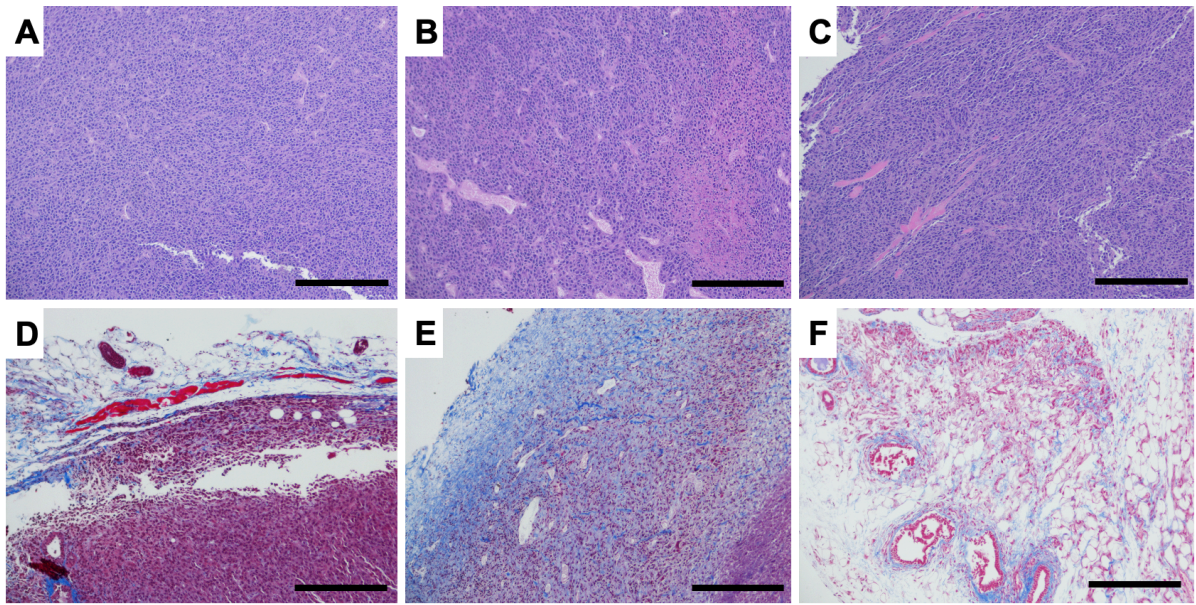

**Figure 5. Histopathological assessment of response to immunotherapy.** Representative microscopic images of (**A-C**) H&E and (**D-F**) Masson's trichrome stained tissue sections from tumors in (**A,D**) CTRL, (**B,E**) αCTLA4, and (**C,F**) αPDL1 treatment groups are presented. The scale bars represent 250 μm.

**Discussion**

The main goal of the current study is to test if Raman spectroscopy is sensitive to the compositional changes in the tumors in response to treatment using immune checkpoint inhibitors. The second goal is to test the specificity of the method to distinguish between the treatment responses mediated by distinct biological mechanisms. Due to their widespread use in preclinical studies of immunotherapy, we employed the CT26 murine colorectal tumor xenografts in these studies to compare their responses to treatment with three doses each of anti-CTLA-4 and anti-PD-L1 antibodies. Unlike other forms of treatment, immunotherapy often does not result in immediate reduction in tumor size due to an observed initial increase in volume mediated by immune cell infiltration. Therefore, we did not observe profound changes in tumor volume in response to αCTLA4 and αPDL1 treatment (**Fig. S3, supporting information**). However, histopathological assessment of the tumors revealed subtle but consistent differences in their overall morphology due to treatment with immune checkpoint inhibitors (**Fig. 5**). By leveraging the ability of Raman spectroscopy to capture biochemical changes in tissue composition in advance of their morphological manifestation, we used multivariate data analysis techniques to unravel the latent information from Raman spectra of tumors treated with immune checkpoint inhibitors. The MCR-ALS analysis provided promising evidence of spectral differences that can be tied to the compositional constituents of the tumors. The subtle but statistically significant differences observed in the scores of MCR loadings resembling lipids, nucleic acids, and collagen showed that the response to immunotherapy using αCTLA4 and αPDL1 manifests in changes in the composition of the tumor microenvironment.

These observations are consistent with the emerging evidence of the role of metabolism and tumor microenvironment in modulating immune responses. The differences in the scores of lipid-like components between the treatment classes can be attributed to the differential lipid metabolism observed in the cellular and extracellular components of the tumor microenvironment in response to immunotherapy.[18,19,41] For example, Lin and coworkers recently observed that the tissue-resident memory T cells rely on fatty acid oxidation for survival and compete with gastric adenocarcinoma cells for lipid uptake.[42] Immunotherapy with PD-L1 blockade enabled the memory T cells to outcompete cancer cells and contribute to antitumor immune response.[42] Similarly, another study in regulatory T cells showed that the inhibition of a member of fatty acid binding proteins (FABP5) that facilitate uptake and intracellular lipid trafficking resulted in mitochondrial changes, impaired lipid metabolism, and reduced proliferation, thus providing evidence that lipid metabolism is critical for the formation of an immunosuppressive microenvironment by regulatory T cells.[43] Lipid metabolism, particularly fatty acid oxidation, is also being actively investigated in other cells that are known to play a role in immunotherapy such as tumor-associated macrophages, natural killer cells and myeloid-derived suppressor cells.[18,19,44-46]

Several studies have shown that the fibrotic tumor microenvironment protects tumors from immune surveillance and promotes resistance to therapy.[20] A study in pancreatic cancer showed that the inhibition of focal adhesion kinase reduced fibrosis in the tumor microenvironment and improved T cell infiltration into tumors.[21] It was observed in preclinical models of metastatic breast cancer reduction of fibrosis through CXCR4 blockade alleviated immunosuppression and significantly improved their response to immunotherapy.[22] Similarly, hypoxia (which correlates with high collagen density) is known to suppress the antitumor immune response, particularly due to PD-L1 upregulation, in a variety of cancers.[47] While most investigations have focused on the study of immunosuppressive effects of increased collagen density, they do not evaluate the changes in the collagen content in sensitive tumors in response to immune checkpoint inhibitor therapy. Our results show that there is a significant decrease in the collagen-like MCR scores due to $\alpha$CTLA4 treatment. However, a similar decrease in the scores of the same component due to $\alpha$PDL1 was not statistically significant at the chosen threshold. We hypothesize that the observed decrease in the collagen content of the tumors after immunotherapy may be a result of extracellular matrix remodeling due to immune cell infiltration and tumor cell kill due to immune checkpoint blockade. The results of our MCR-ALS analysis and the subtle differences in the spatial heterogeneity of the component scores introduced by the immunotherapy provide important starting points for detailed investigations to understand the underlying biological processes.

Our binary SVM based supervised classification routine by employing a leave-one-mouse-out protocol showed that a combination of Raman spectroscopy and machine learning can potentially provide an automated route for identification of response to immunotherapy. The leave-one-mouse-out protocol allows us to completely avoid the representation of spectra from test mice in the training dataset and prevent overfitting of the classifier. Furthermore, the impressive accuracy of the SVM comparison between the $\alpha$CTLA4 and $\alpha$PDL1 groups shows that the two immunotherapy strategies result in distinct spectral responses, consistent with the significantly different mechanisms underlying their antitumor immune responses.[5] This observation was also in agreement with our random forest analysis that revealed two separate sets of spectral markers with little overlap that most strongly contribute to the prediction of response to each immunotherapeutic strategy. While the peak assignments for these predictors (**Table ST2**, supporting information) do not readily allow the identification of independent molecular constituents, they show that multivariate analysis of Raman spectra can distinguish between treatment responses guided by distinct underlying biological mechanisms.

Taken together, our results demonstrated that optical methods are capable of non-invasive assessment of tumor response to immunotherapy. The multivariate data analysis of spectra allowed us to overcome the limitations posed by the low throughput of the method, particularly in a challenging application traditionally characterized by subtle and slow molecular changes. We expect that the promising results obtained in this pilot study coupled, with the ability to readily adapt the developed experimental and computational frameworks to new applications, will catalyze new optical spectroscopic investigations of poorly understood aspects of immunotherapy. Future studies aimed at the prediction of tumor responses to immunotherapy using single or multiple agents and their combination with chemotherapy and radiation will pave the way for clinical translation of Raman spectroscopy for treatment planning and monitoring applications.

**Material and Methods**

*Cell culture and Tumor xenografts*

The CT26 colon carcinoma cells were purchased from the American Type Culture Collection (Product: ATCC CRL-2638, Lot Number: 58494154) and passaged according to established protocols. The cells from third and fourth passages were used for tumor inoculation. Cells were injected into the hind flanks of 6-8-weeks-old Balb/c mice (n = 25) purchased from Jackson Laboratories. The mice were maintained at the University of Arkansas Central Laboratory Animal Facility with 12 h light/dark cycles and provided ad libitum access to food and water. All experiments were approved by the Institutional Animal Care and Use Committee at the University of Arkansas (IACUC protocol 18065). Mice were randomly assigned to one of three treatment groups - mouse IgG isotype (control, n = 10), anti-mouse CTLA-4 ($\alpha$CTLA4, n = 8), and anti-mouse PD-L1 ($\alpha$PDL1, n = 7).

*Immunotherapy Agents*

The immunotherapy agents were purchased from BioXcell (Control: MPC-11, $\alpha$CTLA4: UC10-4F10-11, $\alpha$PDL1: 10F.9G2). Once tumor volumes reached between 80-120 mm$^3$, mice were treated with 3 doses of 100 μg anti-CTLA-4 or 200 μg anti-PD-L1, or 100 μg of IgG control dissolved in 100 μL of sterile saline via intraperitoneal injection. These doses were delivered on Days 1, 4, and 7 with Day 1 indicating the first day of treatment. Animals were euthanized 3 days after their last dosage. Tumors were excised and snap-frozen for Raman spectroscopy and histopathology.

*Raman spectroscopy*

We performed Raman spectroscopic measurements on the tumors according to previously established protocols. The snap-frozen tumor samples were thawed in phosphate-buffered saline (PBS) and fixed in 10% neutral buffered formalin. The fixed tumors were washed thoroughly in PBS, flattened, and sandwiched between a glass coverslip and an aluminum block with the regular addition of PBS to prevent dehydration during Raman measurements. The tumors were mapped using a fiber-optic probe mounted on a motorized translational stage (T-LS13M, Zaber Technologies Inc., travel range: 13 mm).[48] The lensed fiber-optic probe (Emvision LLC, probe diameter: 2 mm) comprised of separate optical fibers for laser delivery and collection of backscattered Raman photons to provide an estimated tissue sampling volume of 1 mm$^3$. The probe was interfaced with a custom-built portable clinical Raman system that consists an 830 nm diode laser (Process Instruments, maximum power: 500 mW), a spectrograph (Holospec f/1.8i, Kaiser Optical Systems), and a thermoelectrically (TE)-cooled CCD camera (PIXIS 400BR, Princeton Instruments).[48] Using the quartz coverslip to flatten the tumor and maintaining a uniform gap between the probe and the coverslip, the laser power was maintained at approximately 20 mW. The spectra were acquired from spatially distinct locations on both sides of the flattened tumor by raster scanning along a 2D rectangular grid of points spaced approximately 1 mm apart using a LabVIEW interface. We acquired a total of approximately 7500 spectra from the 25 tumors in treatment and control groups using 5 seconds (10 accumulations of 0.5 seconds each to prevent CCD saturation) acquisition time for each spectrum.

*Data analysis*

The analysis of Raman data was performed using in-house code developed in MATLAB 2017b (Mathworks) software. We calibrated the wavenumber axis using acetaminophen and restricted our analysis to the fingerprint region between 600 cm$^{-1}$ and 1800 cm$^{-1}$ in this study. The acquired Raman spectra were pre-processed using fifth-order polynomial-based background subtraction, median filtering, and vector normalization (Euclidean norm of each spectrum is set to unity) to alleviate the contributions of fluorescence, cosmic rays, and laser power variations, respectively.[25] The pre-processing steps preserved the size of the spectral dataset as the employed transformations did not involve any spatial averaging or dimensionality reduction steps.

Next, we used multivariate curve resolution – alternating least squares (MCR-ALS) analysis to decompose the spectral matrix into a small set of tissue constituent-like pure component spectra (loadings) and their contributions (scores) to each spectrum in the dataset under a non-negativity constraint on both the resultant matrices.[39] The imposition of spectral equal length constraint allowed us to use the MCR score for each loading as a measure of the abundance of the pure component it resembles and compare the scores across the treatment groups. We plotted the distribution of scores of key MCR loadings that resembled biological constituent spectra using violin plots with embedded box and whisker plots.[49] The few outliers beyond the whiskers were not shown for clarity and optimal plotting. The statistical significance of differences in the medians of MCR score distributions across the classes was assessed using a two-sided Wilcoxon rank-sum test statistic with a conventional threshold ($p < 0.05$ to consider medians different). We also quantified the effect size of the differences across classes using the Wendt formula for rank biserial correlation.[50] The MCR score maps were used to determine spatial heterogeneity using the method described by Brooks et al. for medical images.[51] The proposed index of heterogeneity quantifies spatial heterogeneity based on the smoothness of intensity transitions in the images (**Fig S2**, supporting information). For each 4 x 4-pixel sub-image within an MCR score map, the deviation of intensity transitions from the smoothest possible transition is calculated for all possible pixel pairs and integrated over normalized length to obtain heterogeneity index. The heterogeneity indices thus obtained for all sub-images from MCR score maps obtained from tumors in all the treatment groups are plotted using box and whisker plots to visualize differences in spatial heterogeneity induced by immune checkpoint inhibition therapy.

We used support vector machines (SVM) based classification routine for the automated detection of response to immunotherapy and distinguishing between the response of different immunotherapeutic agents. SVM employs constrained quadratic optimization to derive class boundaries in higher dimensional spaces for nonlinear classification problems.[52] We used the LIBSVM library to perform C-SVM classification using a radial basis kernel for nonlinear mapping between the spectra and their class labels.[53] The optimal C-SVM parameters (cost and kernel parameter gamma) were found by a grid search over a wide range. To avoid the spectral representation of tumors from the test dataset in the training dataset, we adopted a leave-one-mouse-out approach for our SVM analysis. This involves iteratively excluding all the spectra belonging to each mouse tumor from the training dataset and subjecting them as an independent test dataset to the SVM models trained on the remaining spectral dataset. To avoid skewing the SVM predictions due to varying class sizes across the treatment groups, we performed 20 iterations of the leave-one-mouse-out analysis and considered the median accuracy values for each mouse. The predicted class labels obtained by testing all the spatially distinct spectra from every mouse allow us to assign an overall class prediction for each animal by thresholding on prediction frequencies. For the binary classification in the study, we assigned an animal to the majority class only if its membership was at least 30% higher than the random chance, i.e. over 65%. By extension, the mice were misclassified as the incorrect class only if fewer than 35% spectra are correctly classified. All other mice, where the rate of correct classification lies between 35 and 65% were considered unclassified.

We trained bagged random forest classifiers using the TreeBagger class in MATLAB to quantify the out-of-bag error of the spectral prediction between the control and treatment classes. We trained 100 decision trees in the forest to track the evolution of the out-of-bag error with the increase in the number of trees and obtain measures of permutation predictor importance for all the Raman spectral features.[54] We next ranked the features in the decreasing order of their importance to prediction and identified the top 5 non-neighbor features for each comparison to understand the similarities and differences between them.

*Histopathology*

The formalin-fixed tumors were embedded in paraffin and sectioned on to glass slides for histopathology after the acquisition of Raman spectra. The sections were used to perform hematoxylin and eosin (H&E) staining to observe tumor morphology and Masson's trichrome staining for collagen using standard protocols by the Oncology Histology Core at Johns Hopkins University. The slides were imaged using a Nikon fluorescence microscope using a 10X objective.

**Acknowledgments**

S.K.P. acknowledges the support of the SLAS Graduate Education Fellowship Grant. N.R. acknowledges support from the National Cancer Institute (R01CA238025). I.B. acknowledges support from the National Cancer Institute (R01 CA238025), the National Institute of Biomedical Imaging and Bioengineering (2-P41-EB015871-31), and the National Institute of General Medical Sciences (DP2GM128198). The schematic in Figure 1 was partially created with BioRender.com.

**Supporting Information**

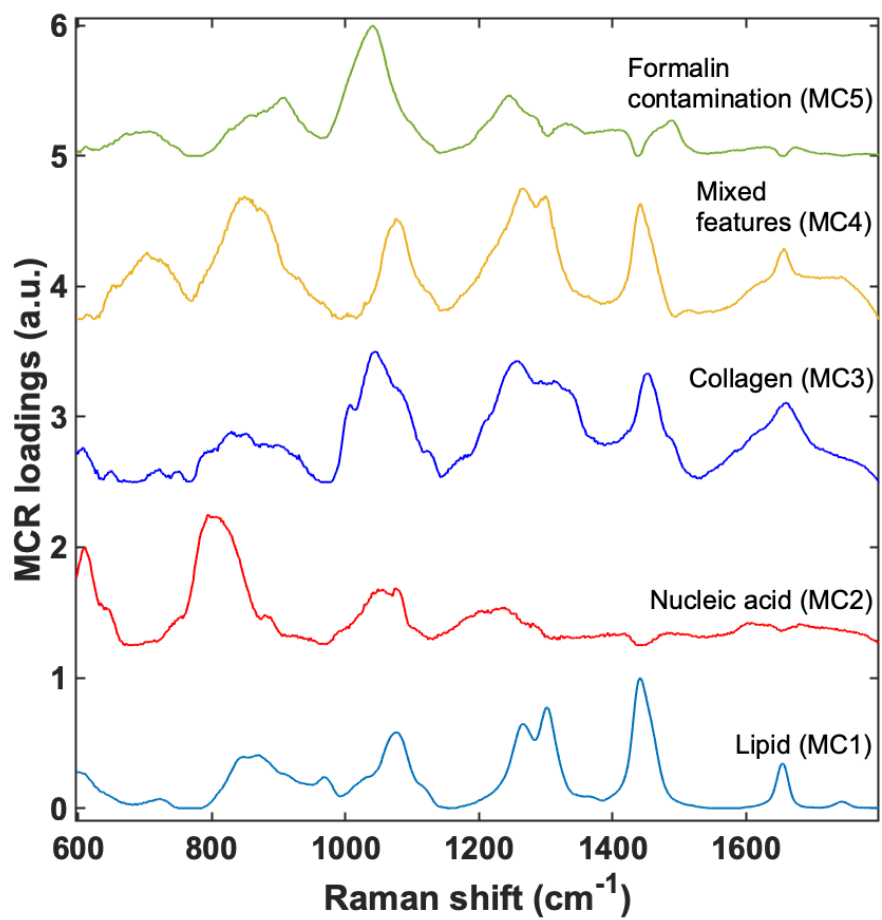

**Figure S1.** The complete set of component loadings derived from MCR-ALS analysis are provided. The biological assignments are marked for each component.

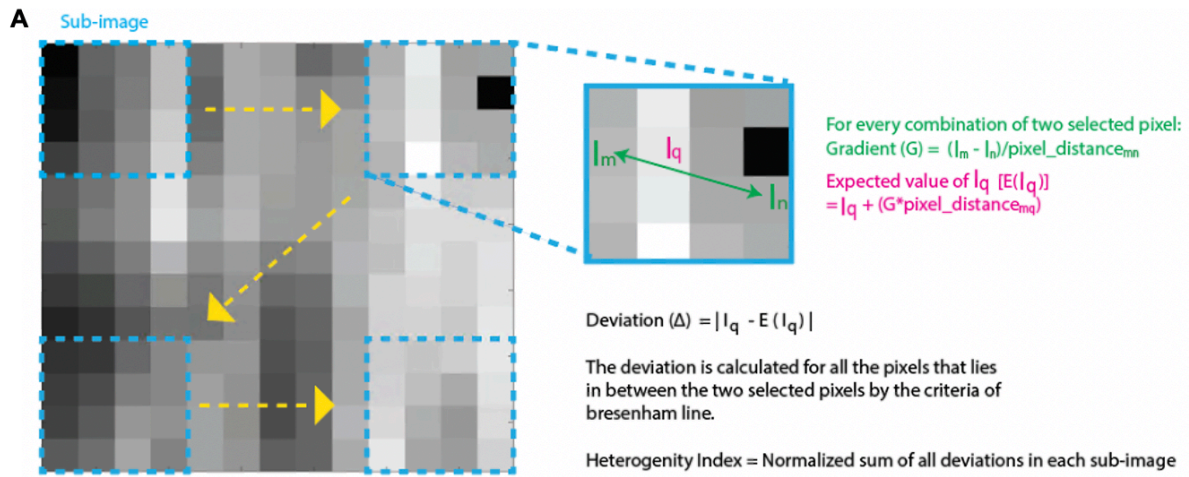

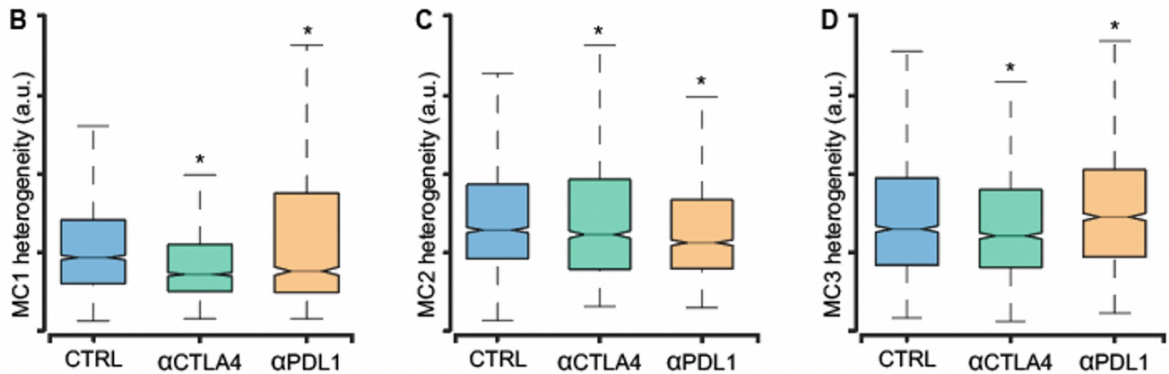

**Figure S2. Raman assessment of changes in spatial heterogeneity in response to immunotherapy.** (**A**) A schematic of the method used for the measurement of heterogeneity index is shown. Box and whisker plots showing the median and interquartile range for the spatial heterogeneity indices derived from the MCR score maps of the tumors in the treatment groups are presented for (**B**) MC1, (**C**) MC2 and (**D**) MC3. The statistical significance of the differences in the heterogeneity of the immune checkpoint inhibitor-treated tumors compared to the controls are assessed by Wilcoxon rank sum test. * indicates p-value < 0.05.

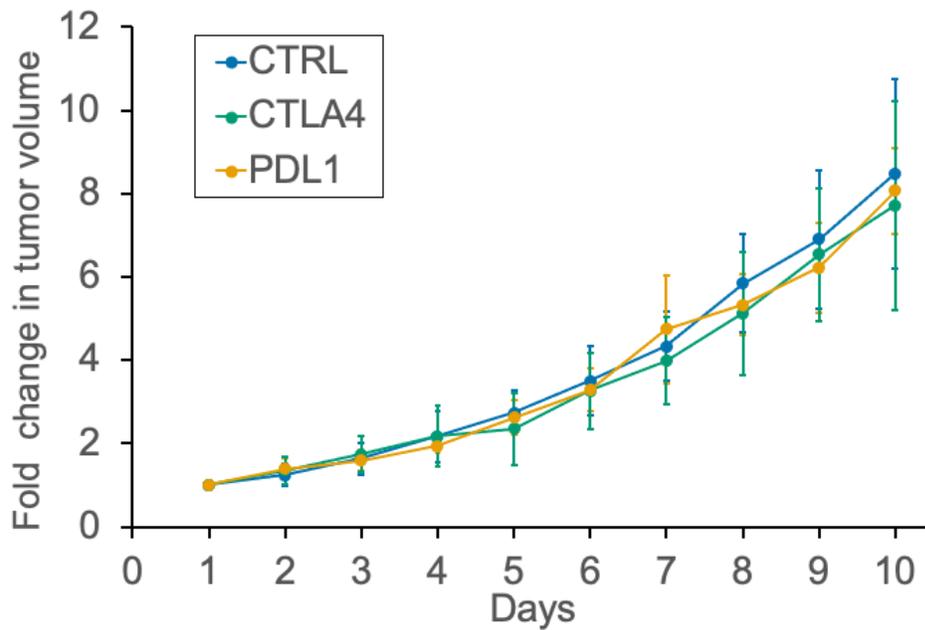

**Figure S3. Evolution of tumor volume with immunotherapy.** The curves show the mean (marker) and standard deviation (error bars) of the volume (fold change compared to day 1) for tumors in the CTRL, αCTLA4, and αPDL1 treatment groups show mean and standard deviation of the volumes (fold change compared to day 1).

**Table ST1.** Table of MCR-ALS component spectral peak assignments

| Observed Raman peaks in the MCR loadings (cm$^{-1}$) | | | | | Raman band assignment from literature |
|---|---|---|---|---|---|
| **MC1** | **MC2** | **MC3** | **MC4** | **MC5** | |
| | 793 | | | | O-P-O stretching in DNA |
| | 813 | | | | O-P-O stretching in DNA and RNA |
| | | 851 | 849 | | C-C stretch of proline in collagen |
| | | | | 908 | Formalin contamination during tissue fixation |
| | | 930 | 930 | | C-C vibration in collagen backbone |
| | | 1042 | | | Proline in collagen |
| | | | | 1042 | Formalin contamination during tissue fixation |
| 1078 | | | | | C-C stretch |
| | 1073 | | | | PO$_2^-$ symmetric stretching in DNA |
| | 1235 | | | | PO$_2^-$ asymmetric stretching in DNA |
| | | | | 1247 | Formalin contamination during tissue fixation |
| | | 1256 | | | Amide III in collagen |
| | | | 1262 | | Amide III in collagen |
| 1266 | | | | | CH$_2$ in-plane deformation (Triglyceride) |
| 1302 | | | 1301 | | CH vibration (Triglyceride) |
| | | 1315 | | | CH$_3$CH$_2$ twisting modes of collagen |
| | | | | | CH$_3$CH$_2$ wagging modes of collagen and nucleic acids |
| 1442 | | | 1442 | | CH$_2$ bending mode (Triglyceride) |
| | | 1448 | | | CH$_2$ bending mode in collagen |
| | | | | 1489 | Formalin contamination during tissue fixation |
| 1654 | | | | | C=C lipid stretch |
| | | 1657 | 1657 | | α-helical structure of amide I in collagen |

**Table ST2.** Assignment assignment for the top spectral predictors derived from random forest analysis

| Observed Raman peaks | Raman band assignment from literature |
|---|---|
| 981 | C-C stretching (proteins) |
| 1065 | =CH bending (lipids) |
| 1323 | C-C stretch (lipids) |
| 1386 | $\delta CH_3$ symmetric band (lipid) |
| 1574 | Nucleic acid modes |
| 1596 | C=C in-plane bending mode (phenylalanine) |
| 1652 | Amide I (collagen) / C=C stretch (lipids) |
| 1763 | C=0 stretch (lipids) |